Static and dynamic measures of human brain connectivity predict complementary aspects of human cognitive performance

Aurora I. Ramos-Nuñez[a], Simon Fischer-Baum[a], Randi C. Martin[a], Qiuhai Yue[a], Fengdan Ye[b,d], and Michael W. Deem[b,c,d]

[a] Department of Psychology, Rice University, Houston, TX

[b] Department of Physics & Astronomy, Rice University, Houston, TX

[c] Department of Bioengineering, Rice University, Houston, TX

[d] Center for Theoretical Biological Physics, Rice University, Houston, TX

Corresponding author
Randi C. Martin, PhD
Department of Psychology, MS-25
Rice University
P. O. Box 1892
Houston, TX 77251

## Abstract

In cognitive network neuroscience, the connectivity and community structure of the brain network is related to measures of cognitive performance, like attention and memory. Research in this emerging discipline has largely focused on two measures of connectivity – modularity and flexibility – which, for the most part, have been examined in isolation. The current project investigates the relationship between these two measures of connectivity and how they make separable contribution to predicting individual differences in performance on cognitive tasks. Using resting state fMRI data from 52 young adults, we show that flexibility and modularity are highly negatively correlated. We use a Brodmann parcellation of the fMRI data and a sliding window approach for calculation of the flexibility. We also demonstrate that flexibility and modularity make unique contributions to explain task performance, with a clear result showing that modularity, not flexibility, predicts performance for simple tasks and that flexibility plays a greater role predicting performance on complex tasks that require cognitive control and executive functioning. The theory and results presented here allow for stronger links between measures of brain network connectivity and cognitive processes.

Research in cognitive neuroscience has typically focused on identifying the function of individual brain regions. Recent advances, however, have led to thinking about the brain as consisting of interacting subnetworks that can be identified by examining connectivity across the whole brain. This emerging discipline of cognitive network neuroscience has been made possible by combining methods from functional neuroimaging and network science (Bullmore et al., 2009; Medaglia et al., 2015; Mill et al., 2017; Sporns, 2014). Functional and diffusion MRI methods provide a rich source of data for characterizing the connections – either functionally or structurally – between different brain regions. Using these data, network science provides mathematical tools for investigating the structure of the brain network, with brain regions serving as nodes, and the connections between brain regions serving as edges in the analysis.

Under this framework, the structure of the brain network can be characterized with a variety of measures. For example, one measure, network modularity, captures the extent to which a network has community structure, by dividing the brain into different modules that are more internally dense than would be expected by random connections (Newman, 2006). A different measure, network flexibility, characterizes how frequently regions of the brain switch allegiance from one module to another, over time (Bassett et al., 2011). Going forward, a major challenge of cognitive network neuroscience is to determine the relationship between measures of brain network structure and cognitive processes (Sporns, 2014).

Relating individual differences in brain network structure to behavioral performance on cognitive tasks provides one tool for addressing this challenge. Both modularity and flexibility have been shown to correlate with variation in cognitive performance. Previous empirical research from our laboratory (Yue et al., 2017) and others (Cohen and D'Esposito, 2016) suggest an interaction between measures of network structure and performance on simple versus complex tasks. For example, previous studies have shown that individual differences in modularity correlate with variation in memory capacity (e.g. Meunier et al., 2014; Stevens et al., 2012). This work has been extended in several recent studies (Yue et al., 2017; Cohen and D'Esposito, 2016), which report a systematic relationship between an individual's performance on a range of behavioral tasks. Both studies report a cross-over interaction; individuals with lower network modularity perform better on complex tasks, like working memory tasks (n-back or operation span), that likely require communication across different brain networks, while individuals with higher network modularity perform better on simpler tasks, like reaction to exogenous cues of attention, simple visual change detection, or low level motor learning. This interaction is predicted by theoretical work on modularity in biological systems (Deem, 2013) which finds that at short time scales, systems with higher modularity afford greater fitness than systems with lower modularity, while at longer times scales, systems with lower modularity are preferred. At the same time, many recent studies have reported that individual variation in network flexibility can explain a host of performance measures, skill learning (e.g. Bassett et al., 2013, 2011), cognitive control (e.g. Alavash et al., 2015; Braun et al., 2015), and mood (Betzel et al., 2016). Indeed, brain network flexibility has been identified as a biomarker of the cognitive construct of cognitive flexibility (Braun et al., 2015).

These prior investigations have focused on either modularity or flexibility as a network measure. Some of the studies have investigated quite different cognitive processes for modularity (e.g., attentional control) and flexibility (e.g., mood). On the other hand, other studies have focused on similar constructs (e.g., working memory for modularity and cognitive control for flexibility or motor learning for both). Thus, these studies leave open the question of the extent to which modularity and flexibility underlie different or similar cognitive abilities. There is an intuitive basis for thinking that they reflect different capacities, as flexibility relates to how much brain networks change over time, and modularity relates to differences in interconnectivity. However, such a conclusion would be premature, since each of these previous studies measured modularity and flexibility in isolation, without considering whether the other measure could also explain variation in the same cognitive performance and whether each contributes independently when the contributions of both are considered simultaneously. No study has directly addressed the basic question of the relationship between modularity and flexibility.[1] This relationship might be one key to understanding how different measures from network neuroscience relate to different cognitive functions.

The current study investigated the relationship between flexibility and modularity, demonstrating a strong relationship between the two measures and presenting a theoretical framework that explains this relationship. Despite this correlation, we argue that modularity and flexibility still reflect distinct cognitive abilities. Specifically, previous theory suggests that high modularity should result in better performance on simple tasks while low modularity should result in better performance on complex tasks (Deem, 2013). This same theory suggests that flexibility should be negatively correlated with performance on simple tasks and positively correlated with performance on complex tasks, and so we investigated the relationship between flexibility, modularity and performance on a battery of simple and complex cognitive tasks. Based on both theory and previous empirical results, we predict that higher brain modularity and/or lower flexibility is related to better performance on simpler tasks while lower brain modularity and/or higher flexibility is related to better performance on more complex tasks.

---

[1] There are several recent studies that have focused on the relationships between static and dynamic measure of connectivity that are tangentially relevant to our current investigation. Thompson and Fransson, (2015) focus on variation in the connectivity between brain regions. They used a sliding time-window of 90 seconds and calculated the correlations coefficient between regions during each time window. They calculated the mean and variance of the connectivity time-series for each subject and pair of connections. Their results showed that the mean and variance of fMRI connectivity time-series scale negatively. Betzel and colleagues (2016) examined variation in connectivity over time, using a sliding window to identify time periods when functional connectivity deviates significantly from the connectivity found for data from the whole session. These time periods with greater deviance showed higher modularity compared with time periods in which the functional connectivity is closer to the average. Both papers suggest some relationship between static and dynamic measures of functional connectivity, though neither investigates the relationship between whole brain flexibility and modularity across individual subjects.

## 2. Methods

### 2.1 Participants

Participants were 52 (18-26 years old, Mean: 19.8 years; 16 males and 36 females) students from Rice University with no neurological or psychiatric disorders. Subjects were given informed consent in accordance with procedures approved by the Rice University Institutional Review Board. Subjects were compensated with $50 upon their participation in both the behavioral and imaging sessions.

### 2.2 Resting-state fMRI

#### 2.2.1 Imaging data acquisition

A high-resolution T1-weighted structural and three resting state functional scans were acquired during a 30-minute session using a 3T Siemens Magnetom Tim Trio scanner equipped with a 12-channel head coil. Scanning was done at the Core for Advanced Magnetic Resonance Imaging (CAMRI) at Baylor College of Medicine. A T1-weighted structural scan was collected first, followed by three consecutive 7-minute functional scans. In between runs, subjects were instructed to remain lying down in the scanner and were informed that the next run would begin shortly. All 52 subjects participated in the imaging session, which lasted about 30 minutes. The T1-weighted structural scan involved the following parameters: TR=2500ms, TE=4.71ms, FoV=256mm, matrix size=256x256, voxel size = 1x1x1 mm[3.]. Functional runs were three 7-minute resting-state scans obtained by using the following sequences: TR = 2000ms, TE = 40ms, FoV = 220 mm, voxel size = 3x3x4 mm, slice thickness = 4mm. A total of 210 volumes per run each with 34 slices were acquired in the axial plane to cover the whole brain.

#### 2.2.2 Preprocessing

Image preprocessing was conducted using afni_proc.py script from example 9a of AFNI_2011_12_21_1014 version software (Cox, 1996) and recommended preprocessing pipelines by Jo et al., 2013) Each functional run was preprocessed separately, including de-spiking of large fluctuations for some time points, slice timing and head motion correction. Then each subject's functional images were aligned to that individual's structural image, warped to the Talairach standard space, and resampled to 3-mm isotropic voxels. Next, the functional images were spatially smoothed with a 4-mm full-width half-maximum Gaussian kernel. A whole brain mask was then generated and applied for all subsequent analysis. A multiple regression model was then applied to each voxel's time series to regress out several nuisance signals, including third-order polynomial baseline trends, six head motion correction parameters and six derivatives of head motion[2]. The outliers censoring process recorded the time points in which the head motion exceeded a distance (Euclidean Norm) of 0.2mm with respect to the previous time point, or in which > 10% of whole brain voxels were considered as outliers by AFNI's 3dToutcount. Then, the recorded time points in each brain voxel were censored by replacing signal at these points using linear interpolation. A multiple regression model was then applied to each voxel's time series to regress out several nuisance signals, including third-order polynomial baseline trends, six head motion correction parameters, and six derivatives of head motion[3]. In order to reduce the effects of low frequency physiological noise and to ensure that no nuisance-related variation was introduced, a bandpass (0.005-0.1 Hz) filtering was conducted in the same regression model (Biswal et al., 1995; Ciric et al., 2016; Cordes et al., 2001; Hallquist et al., 2013). The residual time series after application of the regression model were used for the following network analyses.

---

[2] Some preprocessing pipelines suggest regressing out signal from white matter and ventricles, under the assumption that any signal measured in these regions must be nuisance signal. However, other recent studies have suggested that BOLD response detected in the white matter reflects a local physiological response, which would suggest that regressing out white matter signal is unwarranted (Gawryluk et al., 2014). Aurich et al., (2015) have shown that different preprocessing strategies can have impact on some graph theoretical measures. While there is the concern that white matter activity reflects something other than a nuisance signal, we still went ahead and ran analyses in both ways. It is worth mentioning here that when we regressed out WM and CSF signal, the correlations between flexibility and task performance dropped substantially. We reason that if WM and CSF are simply nuisance variables, then removing them should not remove the relationship between flexibility and the behavioral measures of interest. Therefore, the main results are presented from data where WM and CSF signal were not regressed out and in our supplementary materials in Table 2 we present the data from WM and CSF signal regressed out.

[3] There is an ongoing debate about the best practices for motion correction, which is outside of the scope of the current study (Ciric et al., 2016; Power et al., 2015). Since we are calculating measures of flexibility, censoring out time points and leaving gaps in the data would have not been the best practice. Therefore we replaced censored signal from time points in which the head motion exceeded a distance (Euclidean Norm) of 0.2 mm relative to the previous time point using linear interpolation.

### 2.2.3 Network re-construction, modularity, and flexibility calculation
#### 2.2.3.1 Network re-construction

The whole brain network was re-constructed based on different functional and anatomical brain parcellations including others used in the resting state literature (Craddock et al., 2012; Glasser et al., 2016; Gordon et al., 2016; Power et al., 2011). The full set of results for all parcellation schemes is reported in the supplemental material. The results are largely consistent across parcellation scheme, but were the clearest with the anatomical parcellation from the 84 Brodmann areas (BA) (42 Brodmann areas for left and right hemispheres respectively). Therefore, results from the BA anatomical parcellation are reported below. First, Brodmann area masks were generated using the TT_Daemon standard AFNI atlas (Lancaster et al., 2000), from AFNI_2011_12_21_1014 version. Then the mean time series for each area was extracted by averaging the preprocessed time series across all voxels covered by the corresponding mask. In the network, each Brodmann area served as a node and the edge between any two nodes was defined by the Pearson correlation of the time series for those two nodes. For each subject and each run, edges for all pairs of nodes in the network were estimated, resulting in an 84×84 correlation matrix. While the modularity values varied within subject across the three runs, calculating modularity values from each of the runs separately and then averaging together those three modularity values was highly correlated (r = 0.92, p=1.8*10^(-22) with the modularity value obtained from a correlation matrix that averaged the correlations across the three runs. Thus, the averaged correlation matrix across three runs was later used to calculate modularity for each subject by applying the Newman algorithm (Newman, 2006).

#### 2.2.3.2 Modularity

Modularity is a measure of the excess probability of connections within the modules, relative to what is expected by chance. To calculate modularity, we first took the absolute values of each correlation and set all the diagonal elements of the correlation matrix to zero. Since fewer than .05% of the elements in the matrix were negative and their absolute values were relatively small, taking absolute values did not have a major effect on the results. To show that taking the absolute values did not have major effects on the results, the correlation between modularity and flexibility from raw data is also reported. The resulting matrix was binarized by setting the largest 400 edges (i.e., 11.47% graph density) in the network to 1 and all others to 0. To show that the results were persistent with the cutoff, we also considered 300 (8.61%) and 500 edges (14.34%). The non-linear filtering afforded by the binarization process has been argued to improve detection of modularity by increasing the signal-to-noise ratio (Chen and Deem, 2015). Modularity was defined as in Equation 1 below, where $A_{ij}$ is 1 if there is an edge between Broadmann areas $i$ and $j$ and zero otherwise, the value of $a_i = \Sigma_j A_{ij}$ is the degree of Brodamann area $i$, and e = ½ $\Sigma_i a_i$ is the total number of edges, here set to 300,400 and 500, respectively. Newman's algorthim was applied to the binarized matrix to obtain the (maximal) modularity value and the corresponding partitioning of Brodmann areas into different modules for each subject.

$$M = \frac{1}{2e} \sum_{all\ module\ areas\ i,j} \sum_{within\ this\ module} \left( A_{ij} - \frac{a_i a_j}{2e} \right) \tag{1}$$

### 2.2.3.3 *Flexibility*

Different methods have been proposed for calculating flexibility, either from methods that rely on modeling the networks as a multi-layer ensemble (Bassett et al., 2010; Mucha et al., 2010) (Bassett et al., 2011; Mucha et al., 2010), or from a sliding window approach, in which the scan session is divided into overlapping sub-intervals, modularity is calculated over each sub-interval and each of the parcellation of brain regions into modules is compared across adjacent windows (Hutchison et al., 2013). Here we adopt this latter approach for calculating flexibility, as this sliding window analysis is the most commonly used strategy for examining dynamics in resting-state connectivity (see Hutchison et al., 2013 for a discussion). Flexibility was calculated from the time series data by following a sliding window with 40 time points (Figure 1a). This sliding window of 40 points was chosen because the autocorrelations returned to zero at around time 40 (see Supplemental Information Figure 3) as well as being a length within the norm of previous studies (Leonardi and Van De Ville, 2015; Zalesky and Breakspear, 2015). For each window we obtained a correlation matrix as shown in figure 1B. Figure 1C reflects the computation of $C_i(t)$, the record of which module the ith Brodmann area is in, at time window t, $1 \leq t \leq 165$. Flexibility for a given Brodmann area of a given subject is the number of changes in the value of $C_i(t)$ across the 165 time windows of length 40. Note that the labeling of the modules can change between time points. Therefore, to account for this effect, we relabeled module numbering so that the difference, defined as the number of areas that have $C_i(t+1) \neq C_i(t)$, i = 1-84 is minimized. The assumption behind this is that the allegiance of areas either only have a minimal change, or no change at all because between adjacent windows there is only one time point out of 40 that has changed. This minimized distance is considered the real difference between windows (Figure 1C and 1C). A detailed illustration of the relabeling process is in figure 1E. Flexibility for a given subject is the average of flexibility values from three runs across all Brodmann areas for that subject. The data from the three runs were not concatenated. Such concatenation would introduce a discontinuity in the time series data, include machine artifacts at the beginning of each run, and lead to spurious correlations in the dynamics. Several recent studies have raised questions about the use of this sliding window approach for calculating network flexibility (Hindriks et al., 2016, Kudela et al., 2017). In particular, these studies raise issues about the ability to differentiate signal from noise with small time series of resting-state data. We can address this concern with our data by correlating the flexibility values obtained from each of the three runs across participants. If these flexibility values simply reflect fluctuations caused by noise, then correlations should not be observed across runs. However, flexibility values are significantly correlated with each other ($p < 0.03$, see Supplemental Information), indicating signal has been detected in the data.

### 2.3 Behavioral tasks

Previous empirical research from our laboratory (Yue et al., 2017) and others (Cohen and D'Esposito, 2016) suggests an interaction between measures of network structure and performance on simple versus complex tasks. This interaction is supported by theoretical work on modularity (Deem, 2013) which finds that at short time scales, biological systems with higher modularity are preferred over systems with lower modularity, while at longer times scales, biological systems with lower modularity are preferred. Relating this theory to brain modularity and performance on cognitive tasks, we predict that higher brain modularity would be related to better performance on simpler tasks while lower brain modularity would be related to better performance on more complex tasks. For the purposes of the current research, this simple vs. complex task distinction is operationalized in the following way: complex tasks are those tasks in which executive attention and cognitive control (the ability to ignore preponderant distractors while performing correctly the task at hand) are required to properly perform the task. Simple tasks are those tasks whose performance does not depend on these operations. Because of the engagement of cognitive control, complex tasks typically require longer processing times than simple tasks. The full battery of tasks is described below. Complex tasks in our battery include measures that require shifting between tasks (operation span (Unsworth et al., 2005) and task-shifting), attentional control (visual arrays (Shipstead et al., 2014)), short-term memory maintenance as well as controlled search of memory (digit span (Unsworth and Engle, 2007)), and the resolution component of the ANT, which taps into the coordination of perception and cognitive control. Simple tasks include the alerting and orienting components of the ANT, which measure automatic responses to exogenous attentional cues (Corbetta and Shulman, 2002) and the traffic light task, which is a measure of response to low level visual properties. To limit the number of behavioral measures in the analysis, improve the reliability of the dependent measure, and to tap into the cognitive mechanisms shared by the tasks (Nunnally & Bernstein, 1994; Winer, 1971), two composite scores, simple and complex, were calculated for the 40 subjects who participated in all of the tasks described below. These composites were computed by summing the z-scores (z-scores were calculated by subtracting the mean from each individual score and dividing it by the standard deviation) for the performance measures for the simple and complex tasks. It is important to note that sorting the tasks in a dichotomous manner is merely a practical procedure and not a suggestion that cognitive control cannot be a continuous process as has been suggested in previous work (Rougier et al., 2005). Future work will be to examine cognitive control as a continuous mechanism of varying degrees.

**Operation span.** Subjects were administered the operation span task (Unsworth et al., 2005) to measure their working memory capacity. This task has been shown to have high test-retest reliability, thus providing a stable measure in terms of the rankings of individuals across test sessions (Redick et al., 2012). In this task, for each trial, participants saw an arithmetic problem, e.g., (2×3)+1, and were instructed to solve the arithmetic problem as quickly and accurately as possible. The problem was presented for 2 seconds. Then, a digit, e.g., 7, was presented on the next screen. Subjects judged whether this digit was a correct solution to the previous arithmetic problem by using a mouse to click a "True" or "False" box on the screen. After the arithmetic problem, a letter was presented on the screen for 800ms that subjects were instructed to remember. Then the second arithmetic problem was presented, followed by the digit and then the second letter, with the same processing requirements for both arithmetic problem and letter, and so forth. At the end of each trial, subjects were asked to recall the letters in the same order in which they were presented. The recall screen consisted of a 3×4 matrix of letters on the screen and subjects checked the boxes aside letters to recall. Subjects used the mouse to respond to the arithmetic problem and to recall letters. The experimental trials included set sizes of six or seven arithmetic problem – letter pairs. There were twelve trials for each set size, resulting a total of 156 letters and 156 math problems. The six and seven set size trials were randomly presented.

Before the actual experiment, a practice session was administered to familiarize subjects with the task. The practice session consisted of a block involving only letter recall, e.g., recalling 2 or 3 letters in a trial, a block involving only arithmetic problems, and a mixed block in which the trial had the same procedure as in the experimental trials, i.e., solving the arithmetic problems while memorizing the letters, and recalling them at the end, but with smaller set sizes of 2, 3, or 4. The response times for math problems and accuracy for arithmetic problems and letter recall were recorded. The operation span score is the accuracy for letter recall, calculated as the number of letters that were recalled at the correct position out of total number of the presented letters. The maximum span score is 156.

**Visual arrays task**. A visual arrays task was used to tap visual short-term memory capacity. In this task, subjects were instructed to fixate at the center of the screen. Arrays of 2 to 5 colored squares at different positions on the screen were presented for 500ms, followed by a blank screen for 500ms, and then by multi-colored masks for 500ms. A single probe square was then presented at one of locations where the colored squares had appeared. Subjects had to judge whether the probe square had the same or different color as the one at the same position. The order of different array sizes was random. Each array size condition had 32 trials, half of which were positive response trials and half negative. The visual short-term memory score was calculated by averaging the accuracy across all array sizes.

**Digit span.** In this task a list of numbers were presented in auditory form at the rate of one number per second and participants were required to memorize them. After presenting the last number in the list, a blank screen prompted participants to recall the numbers in the order in which they were presented by typing on the keyboard. Participants were given five trials for each set size starting at two. The program would terminate if participants got fewer than 3 correct trials for that set size (60% accuracy). Digit span was calculated by estimating the list length at which the subject would score 60% correct using linear interpolation between the two set sizes that spanned this threshold.

**Task-shifting task**. In this task, participants responded to an object according to a preceding cue word. The object was either a square or a triangle, and the color of the object was either blue or yellow. If the cue was "color", participants pressed a button to indicate whether the object was blue or yellow, and if the cue was "shape", they pressed a button to indicate whether the object was a square or a triangle. The same buttons were used for the two tasks. The response time was recorded from the onset of the object. For half of the trials, the cue was the same as that in the previous trial, a repeat trial, and for the other half, the cue changed, a switch trial. For each condition, to take into account both response time and accuracy in a single measure, we calculated the inverse efficiency (IE) score (Townsend and Ashby, 1983) (Townsend and Ashby, 1983) defined as mean RT/proportion correct. The task shifting cost was measured as the difference in inverse efficiency score between the repeat and switch trials. We also adjusted cue-stimulus interval (CSI), which is the time between onset of the cue and onset of the object, using CSIs of 200ms, 400ms, 600ms, and 800ms. However, as the effect of modularity on IE did not differ for different CSIs, the data were averaged across CSI.  In total, there were 256 repeat trials and 256 switch trials.

**Attention Network Test.** The Attention Network Test (ANT; Fan et al., 2002) was used to measure three different attentional components: alerting, orienting, and conflict resolution. In this task, subjects responded to the direction of a central arrow, pressing the left or right mouse button to indicate whether it was pointing left or right. The arrow(s) appeared above or below a fixation cross, which was in the center of the screen. The central arrow appeared alone on a third of the trials and was flanked by two arrows on the left and two on the right on the remaining two third of trials.  The flanking arrows were evenly split between a condition in which they pointed in the same direction as the central one, a congruent condition, and a condition in which they pointed in the opposite direction, an incongruent condition. In the neutral condition, there were no flanking arrows. On three-quarters of the trials, the arrow(s) were cued by an asterisk or two asterisks, which appeared for 100ms on the screen. The interval between offset of the cue and onset of the arrow was 400ms. There were four cue conditions: 1) no cue condition, 2), a cue at fixation, 3) double-cue condition, with one cue above and the other below fixation, and 4) spatial-cue condition, where the cue appeared above or below the fixation to indicate where the arrows would appear. Thus, the task had a 4 cue × 3 flanker condition factorial design. The experimental trials consisted of three sessions, with 96 trials in each session, and 8 trials for each condition. For half of all trials, arrows were presented above the fixation and for the other half below. Also, for half of the trials, the middle arrow pointed left and for the other half right. The order of trials in each session was random. Before the experimental trials, 24 practice trials with feedback were given to subjects that included trials of all types.

Response times and accuracy were recorded. Mean RT for each condition for each subject was computed based on correct trials only. As with task shifting, we calculated the inverse efficiency (IE) score for each condition. The alerting effect was computed by subtracting the IE for the no cue condition from the IE for the double cue condition. The orienting effect was computed by subtracting the IE for the center cue condition from the IE for the spatial cue condition. The conflict effect was computed by subtracting the IE for the congruent condition from the IE for the incongruent condition. To make the direction of the conflict effect the same as that of alerting and orienting effects, we reversed the sign of conflict effect. Thus, the more negative the conflict effect value, the greater the interference from the incongruent flankers.

**Traffic light task.** In this task, subjects saw a red square in the center of screen, which was replaced after an unpredictable time delay (from 2 to 3 seconds) by a green circle. Subjects pressed a button as quickly as possible when they saw the green circle. There were 25 trials in total. Mean response time was calculated for each subject.

All fifty-two subjects participated in the operation span and task-shifting tasks. Forty-three of them participated in the ANT task and visual short term memory task, and forty-four subjects participated in the traffic light task and digit span task, as these were done in a different session, and not all subjects returned to participate in all tasks. The interval between neuroimaging and behavioral sessions varied from 0 (i.e., measuring resting-state fMRI and behavior on the same day but during different sessions) to 140 days.

### 2.4 Methods for linking brain and behavior

To limit the number of behavioral measures in the analysis, improve the reliability of the dependent measure, and to tap into the cognitive mechanisms shared by the tasks (Nunnally and Bernstein, 1994)(Nunnally & Bernstein, 1994; Winer, 1971), two composite scores, simple and complex, were calculated for the 40 subjects who participated in all of the tasks described above. These composites were computed by summing the z-scores (z-scores were calculated by subtracting the mean from each individual score and dividing it by the standard deviation) for the performance measures for the simple and complex tasks. A series of planned comparisons were evaluated, drawn from the theoretical literature. Because these tests are all based on *a priori* hypotheses, no correction for multiple comparisons was necessary. The present study examined the relationship between modularity and flexibility, modularity with simple and complex task performance, and flexibility with simple task and complex task performance by calculating Pearson product-moment correlation coefficients. Additionally, in order to measure the unique contribution of modularity and flexibility on simple and complex task performance, partial correlation analyses were used in which the effect of one variable was controlled for to examine the effect of the other.

## 3. Results

### 3.1 Correlations of modularity and flexibility

Using Brodmann areas as nodes and functional connectivity between these nodes (determined from resting state fMRI) as the measure of the strength of edges, we determined modularity and flexibility for each subject. Figure 2a depicts the relationship between modularity and flexibility across our 52 participants. For the 400-edge analysis, modularity values ranged from .33 to .59, with a mean of .47 (standard deviation 0.056) on a scale from 0 to 1.0. Flexibility values ranged from 27 to 43 with a mean of 32.38 (standard deviation 3.40). A strong negative correlation r=-0.78 ($p < 0.001$) was obtained between these two mathematically different measures, modularity and flexibility, which had not been previously reported. The analysis for the 300- and 500-edge yielded negative correlations of r=-0.81 ($p < 0.001$) and r=-0.74 ($p < 0.001$) respectively. In order to show that taking the absolute value does not have a major effect on results, a strong correlation was also found between flexibility and modularity from raw non binarized data (r=-.647, p<.001). Results from other functional and anatomical parcellations including others used in the resting state literature (Craddock et al., 2012; Glasser et al., 2016; Gordon et al., 2016; Power et al., 2011) are reported in Table 1 of supplemental materials.

### 3.2 Consistency of constituent BAs in modules and flexibility across BAs

Previous research has shown variability in brain regions, with some regions exhibiting higher connectivity variability and other regions showing lower variability (Allen et al., 2014), as well as inter-subject variablility in functional connectivity (Mueller et al., 2013). As is shown in Figure 2b, which aggregates flexibility across subjects for each BA, the same pattern was observed in our data set. Some BAs more frequently changed module alignment across time, and therefore had higher flexibility scores, than others. Regions with higher flexibility include the anterior cingulate cortex, ventromedial prefrontal cortex, orbitofrontal cortex, and dorsolateral prefrontal cortex bilaterally, regions typically associated with cognitive control and executive functions. Regions with lower flexibility are those regions involved in motor, gustatory, visual, and auditory processes such as postcentral gyrus, primary motor cortex, primary gustatory cortex, and secondary visual cortex. The regions with higher and lower flexibility from the Brodmann areas anatomical atlas discussed here are similar to those obtained when calculating flexibility from functional atlases (Craddock et al., 2012) (Supplemental Figure 2). In general, the regions showing lower consistency in module assignment, as measured by the average distance of an individual's modular organization to the modular organization of the group-average data (distance is described in Yue et al., 2017) were the same regions showing greater flexibility (r = .606, *p* < .001). One concern for interpreting these analyses is that, Brodmann areas vary in size; therefore it is possible that the heterogeneity in the size of the regions could affect the results. However, in an additional analysis, the size of each BA was partialled out and the correlation between a region's flexibility and its distance from the average modular organization remained, indicating that size did not have an effect on the results.

### 3.3 Relationship with cognitive performance

For task-shifting, there were fifty-two subjects but one performed at chance level of accuracy and one other showed a shift cost more than three standard deviations above the group mean. Thus these two participants were excluded from the analysis. For the visual arrays and ANT task forty-three subjects participated and for the digit span task and the traffic light there were forty-four subjects. Task-shifting results demonstrated that there was a task shifting effect (mean shift cost = 194 ms, SD; t(49) = 14.54 ms, p < 0.0001). The mean accuracy for the visual arrays task was 89.5%, with a standard deviation of 5.4%. This result has been reported previously (Cowan, 2000). The mean capacity for digit span was 4.26, with a standard deviation of 0.94. The results from the ANT task demonstrated a significant alerting effect (mean = 43 ms/proportion correct, SD: 28 ms/proportion correct; t(42) = 10.14, p < 0.001), a significant orienting effect (mean = 49 ms/proportion correct, SD: 29 ms/proportion correct; t(42) = 10.89, p < 0.001), and a significant conflict effect (mean = 139 ms/proportion correct, SD: 42 ms/proportion correct; t(41)= 21.26, p < 0.001), replicating previous findings (Fan et al., 2002). The group mean RT from the traffic light task was 227ms, with a standard deviation of 17ms. All tasks (except for alerting) had medium to high reliabilities. The Spearman-Brown prophecy reliabilities were 0.85 for operation span, 0.84 for visual arrays task, 0.90 for digit span, 0.68 for shifting task, 0.55 for conflict measure of ANT task, 0.38 for orienting measure of ANT task, and 0.94 for the traffic light task. The reliability for the alerting measure of ANT task was -0.31, thus this measure was eliminated from further consideration.

Scores from the seven remaining behavioral tasks were converted into z-scores by subtracting the group mean from the individual score and dividing it by the standard deviation. They were then combined to create a simple composite score that included an orienting measure of the ANT task and the traffic light task and a complex composite score composed of the operation span task, visual arrays task, digit span task, shifting task, and a conflict resolution measure from the ANT task. Specifically, we found that all five complex tasks' measures correlated significantly with the complex composite score ($r = 0.63$, $p < 0.001$ for operation span; $r = 0.49$, $p = 0.001$ for visual arrays; $r = 0.72$, $p < 0.001$ for digit span; $r = 0.58$, $p < 0.001$ for conflict from the ANT task; $r = 0.37$, $p = 0.018$ for shifting), but were not significantly correlated with the simple composite score ($r = -0.25$, $p = 0.13$ for operation span; $r = 0.27$, $p = 0.09$ for visual arrays; $r = 0.08$, $p = 0.63$ for digit span; $r = 0.06$, $p = 0.7$ for conflict from the ANT task; $r = -0.15$, $p = 0.34$ for shifting). For the simple composite score, given that only two measures went into the composite, the correlations between the individual measures and the composite were necessarily high and equivalent ($r = 0.73$, $p < 0.001$ for orienting measures of the ANT task; $r = 0.73$, $p < 0.001$ for the traffic light task). The scores from the simple tasks were not significantly correlated with the complex composite score ($r = 0.03$, $p = 0.84$ for orienting measures of the ANT task; $r = -0.03$, $p = 0.87$ for the traffic light task). The correlation between the simple and complex composite scores was near zero ($r = 0.005$, $p = 0.98$).

A priori correlation analyses revealed a significant negative correlation between modularity measured with 400 edges and the complex composite (r=-0.335, p=0.037). For simpler tasks, individuals with high modularity performed better, with a significant positive correlation between modularity and the simple composite (r=0.403, p=0.011). As might be expected, given the strong negative correlation between modularity and flexibility, there was a significant positive correlation between flexibility measured with 400 edges and the complex composite (r=0.413, p=0.009) and a nonsignificant negative correlation with the simple composite (r=-0.254, p=0.119). The same pattern was observed at different edge densities (see Table 1).

Despite the strong correlation between flexibility and modularity, it is possible that they make independent contributions to explaining individual differences in cognitive performance. As shown in Figure 3 and Table 1, the magnitude of the correlation coefficient between modularity and the simple task composite is larger than the correlation coefficient between flexibility and simple task composite, across edge densities. The opposite pattern is true for the complex tasks. The correlation coefficient between flexibility and task performance is higher than the correlation coefficient between modularity and task performance. This pattern is partly confirmed by partial correlations analysis controlling for the effect of modularity and flexibility on task performance measured in a network with 400 edges to determine the significance of the unique contribution of each. For the simple task composite, the partial correlation for modularity was significant (r=0.40, p=.03), but that for flexibility was not (r=0.13, p=.44). While the partial correlation for the complex task composite and flexibility was not significant (r=0.25, p=.12), it was numerically larger than the coefficient for modularity (r=-0.006, p=.98).

## 4. Discussion

The present results show that the two measures of brain network structure that are treated as independent in the literature – flexibility and modularity – are actually highly related. Still, each seems to make independent contributions to cognitive performance, with modularity contributing more to performance on simple tasks, and flexibility contributing more to performance on complex tasks.

The first major finding in the current study is that two prominent network neuroscience measures – modularity and flexibility – have a strong negative relationship across individuals. In some sense, this finding is consistent with previous studies that have examined relationships between static and dynamic measure of connectivity. For example, Thompson and Fransson (2015) focused on variation in the connectivity between brain regions. They used a sliding time-window of 90 seconds and calculated the correlation coefficients between regions during each time window and then calculated the mean and variance of that coefficient for each connection across all subjects. Connections with a higher mean connectivity tended to have a low variance and vice-versa. Unlike this previous work, the current study focuses on measures that take into account the entire brain network, with modularity a static measure of whole-brain network structure, and flexibility a dynamic measure of whole-brain network structure. Instead of analyzing variation across individual connections, we analyzed variation across individual subjects. We found a negative relationship between static and dynamic measure of functional connectivity.

How might we account for this strong negative correlation between flexibility and modularity? An intuitive explanation for the negative correlation between flexibility and modularity derives from a dynamical systems perspective that views different configurations of brain regions as attractor states, with modularity measuring the depth of the attractor states (Smolensky et al., 1996). Flexibility measures how frequently the brain transitions between states. Deeper states will naturally be more stable and resistant to transitions, leading to a negative correlation between modularity and flexibility. Given the high degree of correlation between these two measures, it is difficult to interpret findings in the literature that report only one of these measures in isolation. In order to test that the relationship between modularity and flexibility is not due to a method-based explanation, we tested this relationship from randomized signal and compared that of the human subjects to that of an artificial network where modularity is matched to the modularity values from the human subjects. Figure 4 shows that the negative correlation coefficient we observed for the 52 human subjects was more than just a method-based correlation.

However, it would be incorrect to conclude that flexibility and modularity are simply two measures of the same property of the brain network. The second major result from the current investigation is that flexibility and modularity make independent contributions to explaining task performance and, therefore, are likely to link to different cognitive processes. Specifically, our results suggest that flexibility may reflect cognitive control processes (Bassett et al., 2013, 2010), while modularity may reflect simple processes like reaction to exogenous cues of attention, simple visual change detection, or low level motor learning. The regions that show the highest flexibility (Figure 2b) are those that have been previously implicated in control and/or multimodal processes. The complex tasks used in the current study all require aspects of control such as switching between tasks, response selection, and maintaining working memory, while the simple tasks do not. Assuming flexibility indexes cognitive control capacity, we can explain why variation in flexibility plays a larger role in explaining performance on complex tasks. Modularity, on the other hand, seems to explain performance for simple processes. That is, network systems that are highly modular favor performance related to simple processes. This result has been shown in previous work from our laboratory (Yue et al., 2017) and in Cohen and D'Esposito (2016). Strong within-module connections favor low level processing while strong between-module connections favor high order processing. However, the unique contribution of modularity and flexibility on task performance needs further work with greater sample sizes.

In contrast to the complex tasks, flexibility is only weakly related to performance on simple tasks, for these simple tasks, there is a clear contribution of modularity on task performance, even as the contribution of flexibility is partialled out. One interpretation of this finding is that simple tasks tend to rely on only a single network module (Yue et al., 2017; Cohen and D'Esposito, 2016). Stronger connections within that module yield better performance on these tasks, while stronger connections between that module and other network modules have a negative impact on performance for these simple tasks. Given that higher modularity scores correspond to higher within module connections and lower between module connections, higher modularity should be related to better performance on simple tasks that rely on a single network module.

However, flexibility has previously been reported to have strong relationships with the ability to learn even in very simple motor tasks (Bassett et al., 2010). How can we account for these seemingly contradictory results? One possibility is that the methods for measuring flexibility differ between the two studies, with the current study using a sliding-window method for calculating flexibility while Bassett and colleagues rely on a multi-layer dynamic modeling approach. There remains open debate about how to appropriately measure network flexibility from resting state data (Hindriks et al., 2016, Kudela et al., 2017). However, it seems unlikely that these methodological differences would flip the direction of the correlation, with individuals who have higher than average flexibility values by one metric consistently showing lower than average flexibility values by the other, or vice versa. A more theoretically interesting possibility is that the initial stages of learning even simple skill benefits from cognitive control operations, making tasks appear more complex. Therefore, at the initial stages of learning, it is beneficial to have a more flexible brain (Bassett et al., 2013). As learning progresses and the task becomes automatized, cognitive control is no longer necessary and the task becomes simpler. Following the theory depicted in Figure 2, as learning progresses, flexibility should decrease and modularity should increase, as has been previously observed (Bassett et al., 2015, 2013).

There are several methodological concerns that arise from the current study. However, we contend that while additional studies should address these limitations, they are unlikely to reverse the two main findings: first, that modularity and flexibility have a strong negative correlation and second, that modularity and flexibility make separable contributions to explaining task performance. One concern is that the negative correlation between modularity and flexibility could have a method-based rationale. In order to ensure this was not the case, we tested the relationship between modularity and flexibility for randomized signal and compared it to that of human subjects as well as to an artificial network where modularity values matched those of the human subjects. As shown in figure 4, the negative correlation coefficient between modularity and flexibility observed in our 52 human subjects was more than just a method-based correlation. A second concern is that there was wide variability in the time that elapsed between collecting the behavioral measures and resting state fMRI data (from 0 to 140 days). Yue et al. (2017) report that the correlations between modularity and task performance weaken with a greater time between behavioral testing and resting state fMRI. Therefore, including longer elapsed times in our analysis did in fact weaken the relationship between either modularity or flexibility, and performance during the individual tasks (Supplemental table 3). However, this concern will have no effect on the relationship between modularity and flexibility, two measures that are calculated over the same set of resting-state fMRI data, and the elapsed time is identical for both flexibility and modularity and the behavioral results. It is unclear how variation in time between behavioral testing and neuroimaging could explain why flexibility has a strong association with complex task performance and modularity has a stronger association with simple task performance. A third concern is that motion artifacts have been known to introduce signal biases in the resting state data and individuals vary in the extent to which they move during scanning (For a review see Ciric et al., 2016). However, our analyses of the relationship between modularity, flexibility and task performance regressed out individual differences in motion (time series with excessive motion) during scanning. Therefore, the results of the current study are not simply an artifact of individual differences in motion. Still, the issue about motion and connectivity relationship is an area that requires more investigation (see Ciric et al., 2016 for a review of recent systematic comparisons of various motion correction practices). In terms of the measures available to study brain network and cognitive function relationships, the current study focused on two, modularity and flexibility, which were appropriate for the questions explored here. Future studies should examine other types of network measures such as the degree of local information integration, global information segregation, local properties such as the degree centrality of a node, or core periphery organization in order to arrive at a complete picture (Medaglia et al., 2014).

A final concern raised by the current study is how the measures of network organization vary as a function of brain parcellation scheme. Specifically, the results reported here focused on an anatomical atlas (Brodmann's areas). Other functionally-defined atlases were used to calculate both modularity and flexibility (see Table 1 of the supplemental material) but were not related to behavioral measures because Yue and colleagues (2017) report the strongest correlations between modularity in the BA parcellation and performance. Similarly, we found that the correlation between modularity and flexibility was significant with all parcellations considered, but highest for the BA parcellations. Other studies have reported low correlations consistency between functional and anatomical atlases. For example, Cohen and D'Esposito (2016) reported that the correlation between modularity from an anatomical parcellation and a simple sequence tapping task was significant but not with modularity from a functional atlas (Cohen & D'Esposito, 2016). Across at least several studies, anatomical atlases appear to be better at characterizing brain networks than functionally defined atlases. Since previous work has argued that inappropriate node definition might mischaracterize brain regions which have distinctive functions (Wig et al., 2011), and found that nodes derived based on task-based fMRI studies did not align well with anatomical parcellations (Power et al., 2011), one might have expected the opposite – that is, that using network nodes determined at least in part from functional activations (e.g., Glasser et al., 2016) would be better for cognitive network neuroscience than anatomically defined atlases. Given the results of the current study and work from other labs, the question of the best parcellation scheme for cognitive network neuroscience remains an open one. However it is beyond the scope of the current study.

## 5. Conclusion

For cognitive network neuroscience to advance, better links between measures of network structure to cognitive and neural computations must be developed (Sporns, 2014). The theory and results presented here, disentangling the effects of two commonly, but interrelated measures, are one step. By considering how different measures of brain structure relate to each other and relate to variation in performance, we can start to develop stronger links between the cognitive and the network sides of this new approach.

**Acknowledgements**

We are grateful for the students who participated in this study.  This work was partially supported by the T.L.L. Temple Foundation. MWD and FDY were partially supported by the Center for Theoretical Biological Physics under NSF grant #PHY-1427654.

**References**

Alavash, M., Hilgetag, C.C., Thiel, C.M., Gie??ing, C., 2015. Persistency and flexibility of complex brain networks underlie dual-task interference. Hum. Brain Mapp. 36, 3542–3562. doi:10.1002/hbm.22861

Allen, E.A., Damaraju, E., Plis, S.M., Erhardt, E.B., Eichele, T., Calhoun, V.D., 2014. Tracking whole-brain connectivity dynamics in the resting state. Cereb. Cortex 24, 663–676. doi:10.1093/cercor/bhs352

Aurich, N.K., Filho, J.O.A., da Silva, A.M.M., Franco, A.R., 2015. Evaluating the reliability of different preprocessing steps to estimate graph theoretical measures in resting state fMRI data. Front. Neurosci. 9. doi:10.3389/fnins.2015.00048

Bassett, D.S., Wymbs, N.F., Porter, M. a., Mucha, P.J., Carlson, J.M., Grafton, S.T., 2010. Dynamic reconfiguration of human brain networks during learning. Proc. Natl. Acad. Sci. 108, 7641. doi:10.1073/pnas.1018985108

Bassett, D.S., Wymbs, N.F., Rombach, M.P., Porter, M.A., Mucha, P.J., Grafton, S.T., 2013. Task-Based Core-Periphery Organization of Human Brain Dynamics. PLoS Comput. Biol. 9, 1–16. doi:10.1371/journal.pcbi.1003171

Bassett, D.S., Yang, M., Wymbs, N.F., Grafton, S.T., 2015. Learning-Induced Autonomy of Sensorimotor Systems. Nat. Neurosci. 18, 744–751. doi:10.1038/nn.3993

Betzel, R.F., Satterthwaite, T.D., Gold, J.I., Bassett, D.S., 2016. A positive mood, a flexible brain 1–15.

Biswal, B., Yetkin, F.Z., Haughton, V.M., Hyde, J.S., 1995. Functional connectivity in the motor cortex of resting human brain using echo-planar MRI. Magn. Reson. Med. 34, 537–41. doi:10.1002/mrm.1910340409

Braun, U., Schäfer, A., Walter, H., Erk, S., Romanczuk-Seiferth, N., Haddad, L., Schweiger, J.I., Grimm, O., Heinz, A., Tost, H., Meyer-Lindenberg, A., Bassett, D.S., 2015. Dynamic reconfiguration of frontal brain networks during executive cognition in humans. Proc. Natl. Acad. Sci. U. S. A. 112, 11678–83. doi:10.1073/pnas.1422487112

Bullmore, E., Bullmore, E., Sporns, O., Sporns, O., 2009. Complex brain networks: graph theoretical analysis of structural and functional systems. Nat Rev Neurosci 10, 186–198. doi:10.1038/nrn2575

Chen, M., Deem, M.W., 2015. Development of modularity in the neural activity of children ' s brains. Phys. Biol. 12, 16009. doi:10.1088/1478-3975/12/1/016009

Ciric, R., Wolf, D.H., Power, J.D., Roalf, D.R., Baum, G., Ruparel, K., Shinohara, R.T., Elliott, M.A., Eickhoff, S.B., Davatzikos, C., Gur, R.C., Gur, R.E., Bassett, D.S., Satterthwaite, T.D., 2016. Benchmarking confound regression strategies for the control of motion artifact in studies of functional connectivity. ArXiv.

Cohen, J.R., D'Esposito, M., 2016. The Segregation and Integration of Distinct Brain Networks and Their Relationship to Cognition. J. Neurosci. 36, 12083–12094. doi:10.1523/JNEUROSCI.2965-15.2016

Corbetta, M., Shulman, G.L., 2002. Control of Goal-Directed and Stimulus-Driven Attention in the Brain. Nat. Rev. Neurosci. 3, 215–229. doi:10.1038/nrn755

Cordes, D., Haughton, V.M., Arfanakis, K., Carew, J.D., Turski, P.A., Moritz, C.H., Quigley, M.A., Meyerand, M.E., 2001. Frequencies contributing to functional connectivity in the cerebral cortex in "resting-state" data. Am. J. Neuroradiol. 22, 1326–1333.

Cowan, N., 2000. The magical number 4 in short-term memory: A reconsideration of mental storage capacity. Behav. Brain Sci. 24, 87–185. doi:10.1017/S0140525X01003922

Cox, R.W., 1996. AFNI: software for analysis and visualization of functional magnetic resonance neuroimages. Comput. Biomed. Res. 29, 162–173. doi:10.1006/cbmr.1996.0014

Craddock, R.C., James, G.A., Holtzheimer, P.E., Hu, X.P., Mayberg, H.S., 2012. A whole brain fMRI atlas generated via spatially constrained spectral clustering. Hum. Brain Mapp. 33, 1914–1928. doi:10.1002/hbm.21333

Deem, M.W., 2013. Statistical Mechanics of Modularity and Horizontal Gene Transfer. Annu. Rev. Condens. Matter Phys. 4, 287–311. doi:10.1146/annurev-conmatphys-030212-184316

Fan, J., McCandliss, B.D., Sommer, T., Raz, A., Posner, M.I., 2002. Testing the efficiency and independence of attentional networks. J. Cogn. Neurosci. 14, 340–7. doi:10.1162/089892902317361886

Gawryluk, J.R., Mazerolle, E.L., D'Arcy, R.C.N., 2014. Does functional MRI detect activation in white matter? A review of emerging evidence, issues, and future directions. Front. Neurosci. doi:10.3389/fnins.2014.00239

Glasser, M.F., Coalson, T.S., Robinson, E.C., Hacker, C.D., Harwell, J., Yacoub, E., Ugurbil, K., Andersson, J., Beckmann, C.F., Jenkinson, M., Smith, S.M., Van Essen, D.C., 2016. A multi-modal parcellation of human cerebral cortex. Nature 536, 171–8. doi:10.1038/nature18933

Gordon, E.M., Laumann, T.O., Adeyemo, B., Huckins, J.F., Kelley, W.M., Petersen, S.E., 2016. Generation and Evaluation of a Cortical Area Parcellation from Resting-State Correlations. Cereb. Cortex 26, 288–303. doi:10.1093/cercor/bhu239

Hallquist, M.N., Hwang, K., Luna, B., 2013. The nuisance of nuisance regression: Spectral misspecification in a common approach to resting-state fMRI preprocessing reintroduces noise and obscures functional connectivity. Neuroimage 82, 208–225. doi:10.1016/j.neuroimage.2013.05.116

Hutchison, R.M., Womelsdorf, T., Allen, E.A., Bandettini, P.A., Calhoun, V.D., Corbetta, M., Della Penna, S., Duyn, J.H., Glover, G.H., Gonzalez-Castillo, J., Handwerker, D.A., Keilholz, S., Kiviniemi, V., Leopold, D.A., de Pasquale, F., Sporns, O., Walter, M., Chang, C., 2013. Dynamic functional connectivity: Promise, issues, and interpretations. Neuroimage 80, 360–378. doi:10.1016/j.neuroimage.2013.05.079

Jo, H.J., Gotts, S.J., Reynolds, R.C., Bandettini, P.A., Martin, A., Cox, R.W., Saad, Z.S., 2013. Effective preprocessing procedures virtually eliminate distance-dependent motion artifacts in resting state FMRI. J. Appl. Math. 2013. doi:10.1155/2013/935154

Lancaster, J.L., Woldorff, M.G., Parsons, L.M., Liotti, M., Freitas, C.S., Rainey, L., Kochunov, P. V., Nickerson, D., Mikiten, S.A., Fox, P.T., 2000. Automated Talairach Atlas labels for functional brain mapping. Hum. Brain Mapp. 10, 120–

131. doi:10.1002/1097-0193(200007)10:3<120::AID-HBM30>3.0.CO;2-8

Leonardi, N., Van De Ville, D., 2015. On spurious and real fluctuations of dynamic functional connectivity during rest. Neuroimage. doi:10.1016/j.neuroimage.2014.09.007

Medaglia, J.D., Lynall, M.-E., Bassett, D.S., 2015. Cognitive network neuroscience. J. Cogn. Neurosci. 27, 1471–91. doi:10.1162/jocn_a_00810

Meunier, D., Fonlupt, P., Saive, A.-L., Plailly, J., Ravel, N., Royet, J.-P., 2014. Modular structure of functional networks in olfactory memory. Neuroimage 95, 264–75. doi:10.1016/j.neuroimage.2014.03.041

Mill, R.D., Ito, T., Cole, M.W., 2017. From connectome to cognition: The search for mechanism in human functional brain networks. Neuroimage 0–1. doi:10.1016/j.neuroimage.2017.01.060

Mucha, P.J., Richardson, T., Macon, K., Porter, M.A., Onnela, J.-P., 2010. Community Structure in Time-Dependent, Multiscale, and Multiplex Networks. Science (80-. ). 328, 876–878. doi:10.1126/science.1184819

Mueller, S., Wang, D., Fox, M.D., Yeo, B.T.T., Sepulcre, J., Sabuncu, M.R., Shafee, R., Lu, J., Liu, H., 2013. Individual Variability in Functional Connectivity Architecture of the Human Brain. Neuron 77, 586–595. doi:10.1016/j.neuron.2012.12.028

Newman, M.E.J., 2006. Modularity and community structure in networks. Proc. Natl. Acad. Sci. U. S. A. 103, 8577–82. doi:10.1073/pnas.0601602103

Nunnally, J., Bernstein, I., 1994. Psychometric Theory, 3rd edn, 1994. McGraw-Hill, New York 3, 701.

Power, J.D., Cohen, A.L., Nelson, S.M., Wig, G.S., Barnes, K.A., Church, J.A., Vogel, A.C., Laumann, T.O., Miezin, F.M., Schlaggar, B.L., Petersen, S.E., 2011. Functional Network Organization of the Human Brain. Neuron 72, 665–678. doi:10.1016/j.neuron.2011.09.006

Power, J.D., Schlaggar, B.L., Petersen, S.E., 2015. Recent progress and outstanding issues in motion correction in resting state fMRI. Neuroimage 105, 536–551. doi:10.1016/j.neuroimage.2014.10.044

Redick, T.S., Broadway, J.M., Meier, M.E., Kuriakose, P.S., Unsworth, N., Kane, M.J., Engle, R.W., 2012. Measuring working memory capacity with automated complex span tasks. Eur. J. Psychol. Assess. 28, 164–171. doi:10.1027/1015-5759/a000123

Rougier, N.P., Noelle, D.C., Braver, T.S., Cohen, J.D., O'Reilly, R.C., 2005. Prefrontal cortex and flexible cognitive control: rules without symbols. Proc. Natl. Acad. Sci. U. S. A. 102, 7338–7343. doi:10.1073/pnas.0502455102

Shipstead, Z., Lindsey, D.R.B., Marshall, R.L., Engle, R.W., 2014. The mechanisms of working memory capacity: Primary memory, secondary memory, and attention control. J. Mem. Lang. 72, 116–141. doi:10.1016/j.jml.2014.01.004

Smolensky, P., Mozer, Michael, C., Rumelhart, D.E., 1996. Mathematical Perspectiv, in: Smolensky, P., Mozer, M.C., Rumelhart, D.E. (Eds.), Mathematical Perspectives on Neural Networks. Lawrence Erlbaum Associates, Inc.

Sporns, O., 2014. Contributions and challenges for network models in cognitive

neuroscience. Nat. Neurosci. 17, 652–660. doi:10.1038/nn.3690

Stevens, A.A., Tappon, S.C., Garg, A., Fair, D.A., 2012. Functional brain network modularity captures inter- and intra-individual variation in working memory capacity. PLoS One 7. doi:10.1371/journal.pone.0030468

Thompson, W.H., Fransson, P., 2015. The mean–variance relationship reveals two possible strategies for dynamic brain connectivity analysis in fMRI. Front. Hum. Neurosci. 9, 1–7. doi:10.3389/fnhum.2015.00398

Townsend, J.T., Ashby, F.G., 1983. Stochastic Modeling of Elementary Psychological Processes, The American Journal of Psychology. doi:10.2307/1422636

Unsworth, N., Engle, R.W., 2007. On the division of short-term and working memory: An examination of simple and complex span and their relation to higher order abilities. Psychol. Bull. 133, 1038–1066. doi:10.1037/0033-2909.133.6.1038

Unsworth, N., Heitz, R.P., Schrock, J.C., Engle, R.W., 2005. An automated version of the operation span task. Behav. Res. Methods 37, 498–505. doi:10.3758/BF03192720

Wig, G.S., Schlaggar, B.L., Petersen, S.E., 2011. Concepts and principles in the analysis of brain networks. Ann. N. Y. Acad. Sci. 1224, 126–146. doi:10.1111/j.1749-6632.2010.05947.x

Zalesky, A., Breakspear, M., 2015. Towards a statistical test for functional connectivity dynamics. Neuroimage. doi:10.1016/j.neuroimage.2015.03.047

Yue, Q., Martin, R., Fischer-Baum, S., Ramos-Nuñez, A., Ye, F., Deem, M. (2017). Brain modularity mediates the relation between task complexity and performance. J. Cogn. Neurosci. 29, 1532–1546.

Table 1

The correlation coefficients for simple and complex tasks at different edge densities

| | | Correlation Coefficient | |
|---|---|---|---|
| Modularity | | Simple Tasks | Complex Tasks |
| | Number of Edges | | |
| | 300 | 0.25 ($p = .121$) | -0.38 ($p = .016$) |
| | 400 | 0.34 ($p = .030$) | -0.26 ($p = .108$) |
| | 500 | 0.34 ($p = .033$) | -0.28 ($p = .076$) |
| Flexibility | | | |
| | 300 | -0.27 (p = .090) | 0.38 ($p = .016$) |
| | 400 | -0.20 (p = .215) | 0.42 ($p = .007$) |
| | 500 | -0.12 (p = .447) | 0.44 ($p = .004$) |

**(1a)**                    Time points

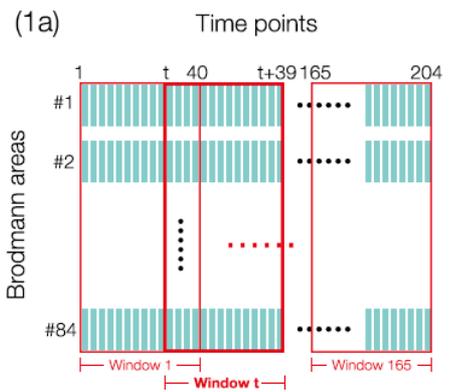

**(1d) Relabel the allegiance of BAs**

| Window t | Window t+1 |
|----------|------------|
| 2 | 4 |
| 2 | 2 |
| 4 | 2 |
| 1 | 1 |
| 3 | 3 |
| ⋮ | ⋮ |
| 4 | 4 |
| 2 | 2 |
| 1 | 1 |

**(2)**     Relabelling process

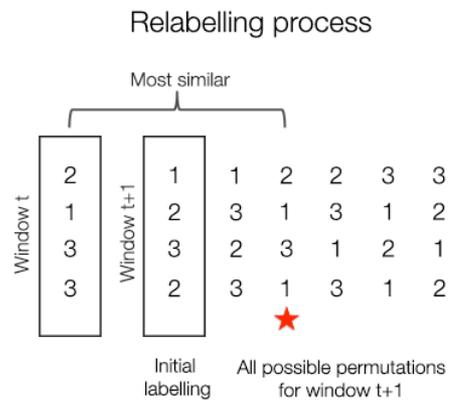

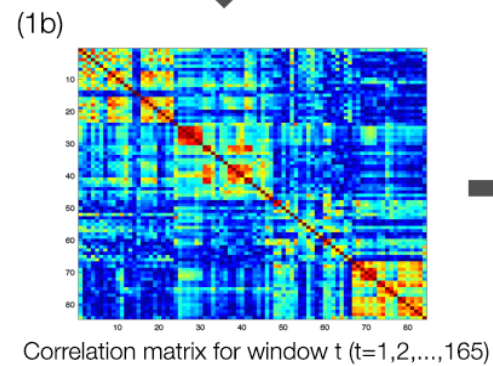

Correlation matrix for window t (t=1,2,...,165)

**(1c)** Allegiance of BAs

| Window t | Window t+1 |
|----------|------------|
| 2 | 1 |
| 2 | 4 |
| 4 | 4 |
| 1 | 2 |
| 3 | 3 |
| ⋮ | ⋮ |
| 4 | 1 |
| 2 | 4 |
| 1 | 2 |

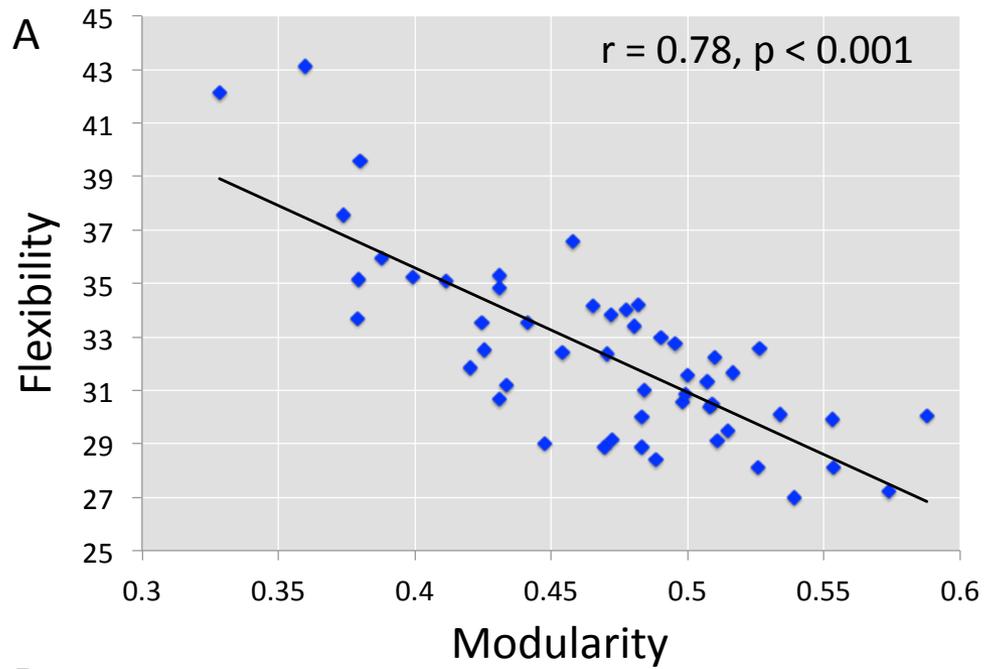

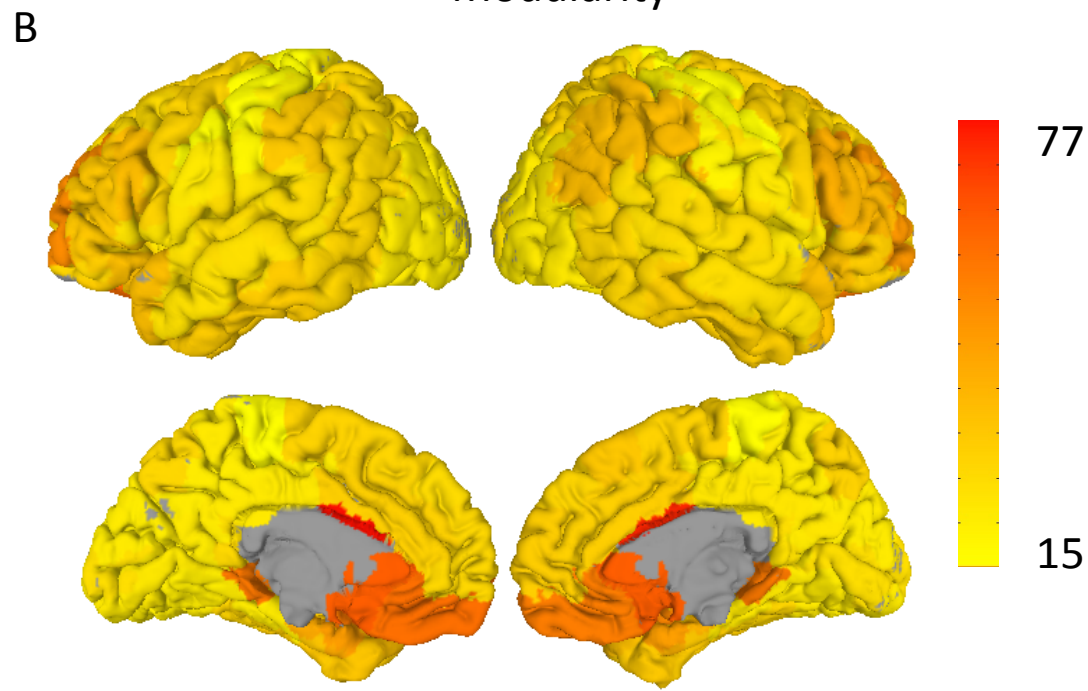

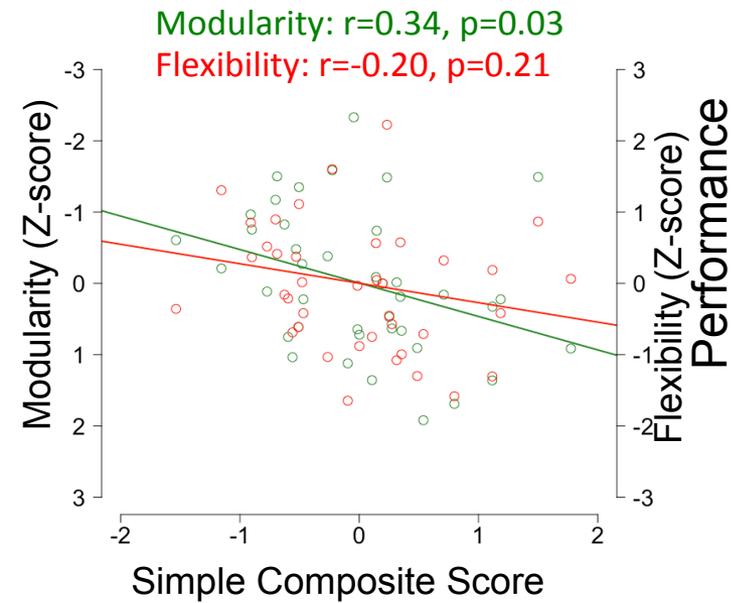

Modularity: r=0.34, p=0.03
Flexibility: r=-0.20, p=0.21

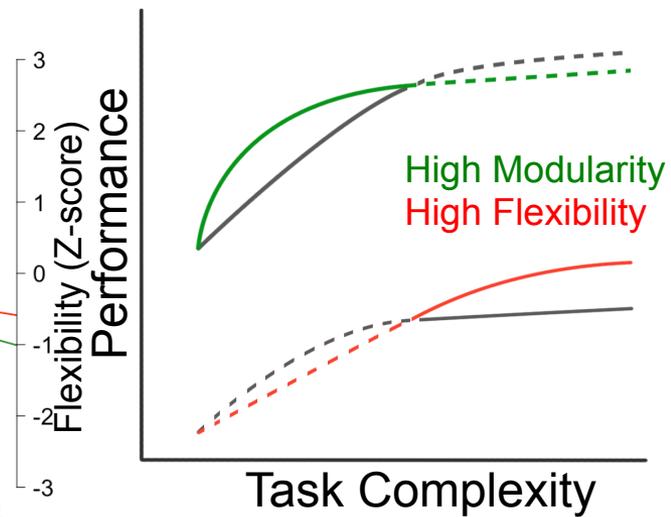

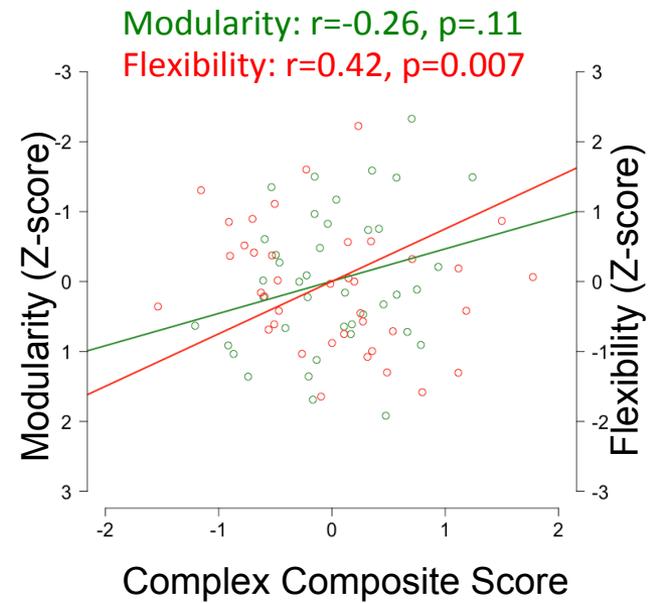

Modularity: r=-0.26, p=.11
Flexibility: r=0.42, p=0.007

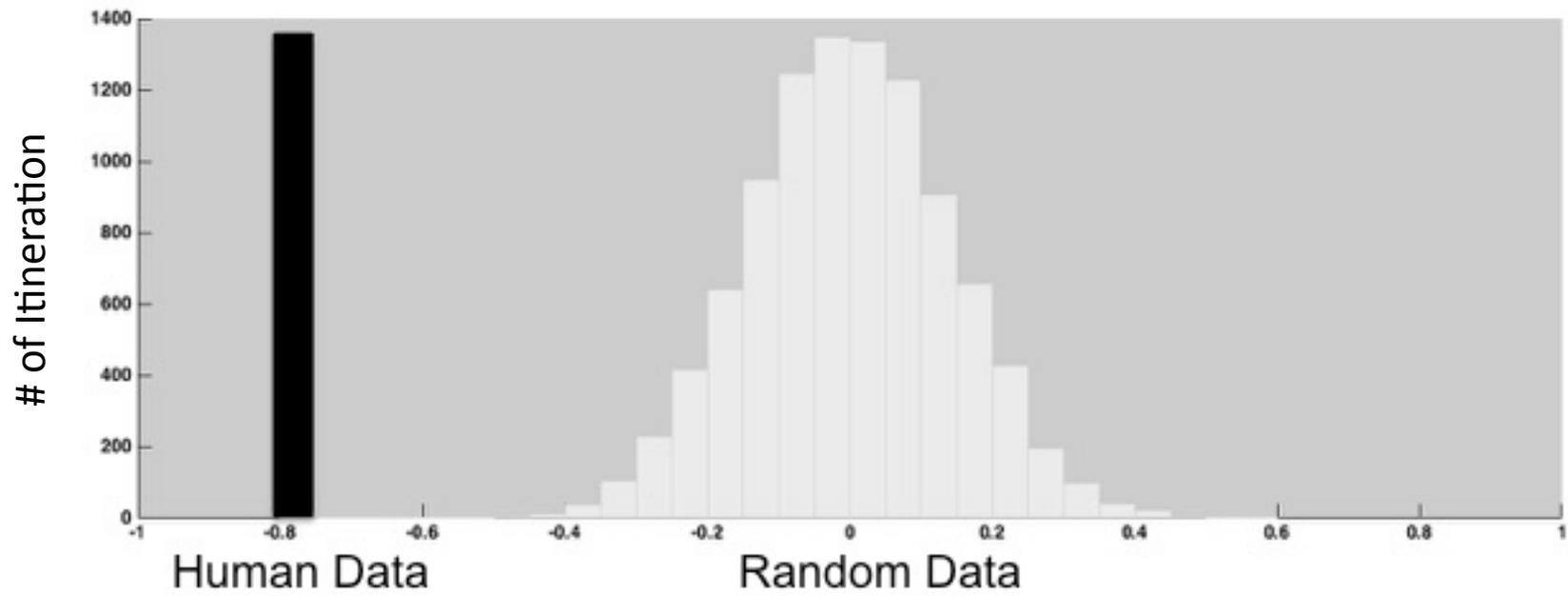

**Supplementary Material**
**Static and dynamic measures of human brain connectivity predict complementary aspects of human cognitive performance**

Aurora I. Ramos-Nuñez[a], Simon Fischer-Baum[a], Randi Martin[a], Qiuhai Yue[a], Fengdan Ye[b,d], and Michael W. Deem[b,c,d]

Corresponding author: rmartin@rice.edu

# 1. Network re-construction

To demonstrate that the correlation between modularity and flexibility persists after using different functional and anatomical parcellations of the brain, the whole brain network was re-constructed based on methods used by others in the resting state literature (Craddock et al., 2012; Glasser et al., 2016; Gordon et al., 2016; Power et al., 2011). As shown in table 1, all the different parcellation methods yielded a negative correlation between modularity and flexibility. However, the magnitude of the coefficient was much larger using the Brodmann areas as the parcellation anatomical map.

# 2. Relationship with task performance

Figure 1 supplemental illustrates the relationship between modularity and task performance and flexibility and task performance during tasks varying from simple to complex processes. Simple tasks include the traffic light and the orienting effect from the Attention Network Task (ANT). Complex tasks involve the Operation Span, Digit Span, Visual Short-term Memory, the Conflict Effect from the ANT, and the Task-Shifting. For the purposes of the current research, this simple vs. complex task distinction is operationalized in the following way: complex tasks are those tasks in which executive attention and cognitive control (the ability to ignore preponderant distractors while performing correctly the task at hand) are required to properly perform the task. Simple tasks are those tasks whose performance does not depend on these operations. Because of the engagement of cognitive control, complex tasks typically require longer processing times than simple tasks. In terms of complex tasks, flexibility generally shows a larger coefficient than modularity. The opposite is true for simple tasks and modularity: modularity presents with a larger coefficient than flexibility.

# 3. Sliding window parameter estimation

We used a sliding window to compute the flexibility values. The use of a sliding window is justified by a decay of correlations in the time series beyond the width of the time window. As show in Supplemental Figure 3, correlations in the time series data decay beyond 80 s. It is for this reason that we use a sliding window of 80s in the computation of the flexibility values. The average correlation among the runs across all subjects is r=0.40 (all three p-values < 0.03).

## 4. Supplementary Figure captions and Tables

**Figure 1 Supplemental**. The relationship between modularity and flexibility with task performance represented by the magnitude of the coefficient between modularity and task performance and flexibility and task performance organized from simple (left) to complex (right). The center of the figure depicts the theoretical prediction relating performance to tasks at different levels of complexity for individuals with high and low modularity (green curve) and flexibility (red curve).

**Figure 2 Supplemental**. A comparison of flexibility across the brain between: A) an anatomical based parcellated atlas with 84 regions such as Brodmann's Areas (BA) and B) a fuctional parcellated atlas with 100 regions from craddock et. al., 2012. The color bar on the right side of the figures represents flexibility values  going from low (15 for BA atlas and 19 for Craddock atlas) to high (77 fir BA atlas and 54 for Craddock atlas).

**Figure 3 Supplemental.** We show the autocorrelation function of the fMRI time series data.  The autocorrelation function of the signal in each of the 84 Brodmann areas in each of the 3 runs for each of the 52 subjects is calculated. Presented are the average of the results over Brodmann areas, runs, and subjects.

Supplementary Table 1
Modularity and Flexibility correlation coefficients from various anatomical and functional atlases

| | r | p-value |
|---|---|---|
| BA_300_edge | -0.81 | p < 0.001 |
| BA_400_edge | -0.78 | p < 0.001 |
| BA_500_edge | -0.74 | p < 0.001 |
| Glasser et al., 2016 | -0.44 | p = 0.001 |
| Gordon et al., 2016 | -0.34 | p = 0.014 |
| Power et al., 2011 | -0.38 | p = 0.005 |
| Craddock et al., 2012 | | |
|     100 parcellations | -0.49 | p < 0.001 |
|     200 parcellations | -0.37 | p = 0.007 |
|     300 parcellations | -0.42 | p < 0.002 |

Supplementary Table 2
Modularity (M), Flexibility (F), and simple and complex task performance
correlation coefficients when white matter and CSF signals were regressed out. We
do not believe the white mattera and CSF signals are purely noise.

|  | r | p-value |
| --- | --- | --- |
| M&F BA_400_edge | -0.70 | 0.001 |
| M BA_400_edge & simple task performance | 0.23 | 0.161 |
| M BA_400_edge & complex task performance | -0.21 | 0.196 |
| F BA_400_edge & simple task performance | -0.07 | 0.646 |
| F BA_400_edge & complex task performance | 0.24 | 0.145 |

Supplementary Table 3
Correlations between individual tasks and modularity and flexibility without controlling for days in between collecting behavioral measures and resting state fMRI data

|  | Modularity | | Flexibility | |
|---|---|---|---|---|
|  | r | p-value | r | p-value |
| Ospan | -0.410 | 0.011 | 0.335 | 0.040 |
| Dspan | -0.180 | 0.279 | 0.326 | 0.046 |
| Conflict | -0.089 | 0.593 | 0.219 | 0.186 |
| VSTM | -0.166 | 0.319 | 0.118 | 0.482 |
| Shifting | -0.026 | 0.875 | -0.130 | 0.438 |
| Orienting | 0.462 | 0.008 | -0.219 | 0.186 |
| Traffic | -0.190 | 0.254 | -0.066 | 0.695 |

Supplementary Table 4
Correlations between individual tasks and modularity and flexibility controlling for number of days in between collecting behavioral measures and resting state fMRI data

|  | Modularity | | Flexibility | |
|---|---|---|---|---|
|  | r | p-value | r | p-value |
| Ospan | -0.432 | 0.008 | 0.330 | 0.046 |
| Dspan | -0.196 | 0.245 | 0.288 | 0.084 |
| Conflict | -0.090 | 0.595 | 0.224 | 0.182 |
| VSTM | -0.181 | 0.284 | 0.068 | 0.690 |
| Shifting | -0.032 | 0.851 | -0.127 | 0.453 |
| Orienting | 0.431 | 0.008 | -0.210 | 0.213 |
| Traffic | -0.186 | 0.272 | -0.223 | 0.184 |

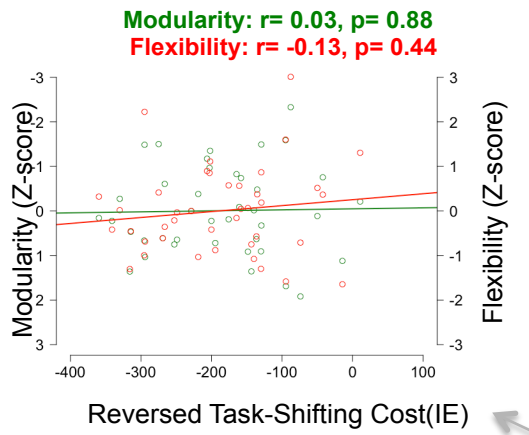

**Modularity: r= 0.03, p= 0.88**
**Flexibility: r= -0.13, p= 0.44**

Reversed Task-Shifting Cost(IE)

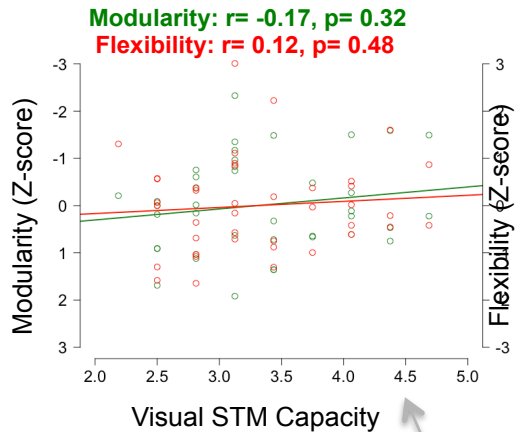

**Modularity: r= -0.17, p= 0.32**
**Flexibility: r= 0.12, p= 0.48**

Visual STM Capacity

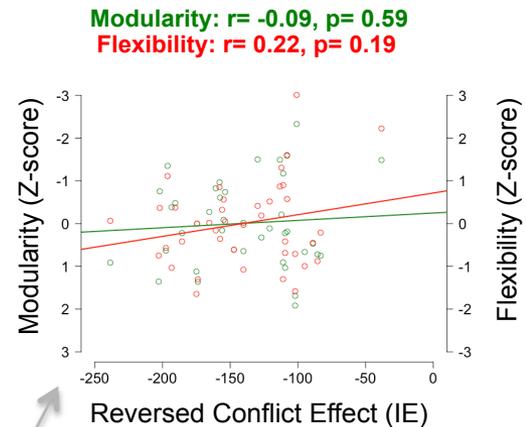

**Modularity: r= -0.09, p= 0.59**
**Flexibility: r= 0.22, p= 0.19**

Reversed Conflict Effect (IE)

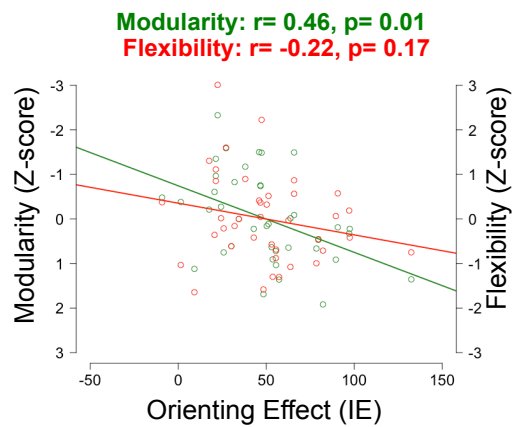

**Modularity: r= 0.46, p= 0.01**
**Flexibility: r= -0.22, p= 0.17**

Orienting Effect (IE)

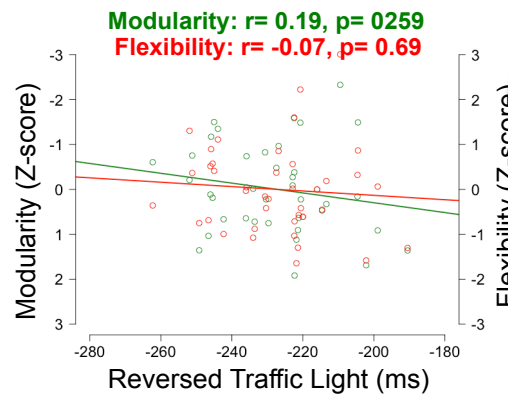

**Modularity: r= 0.19, p= 0259**
**Flexibility: r= -0.07, p= 0.69**

Reversed Traffic Light (ms)

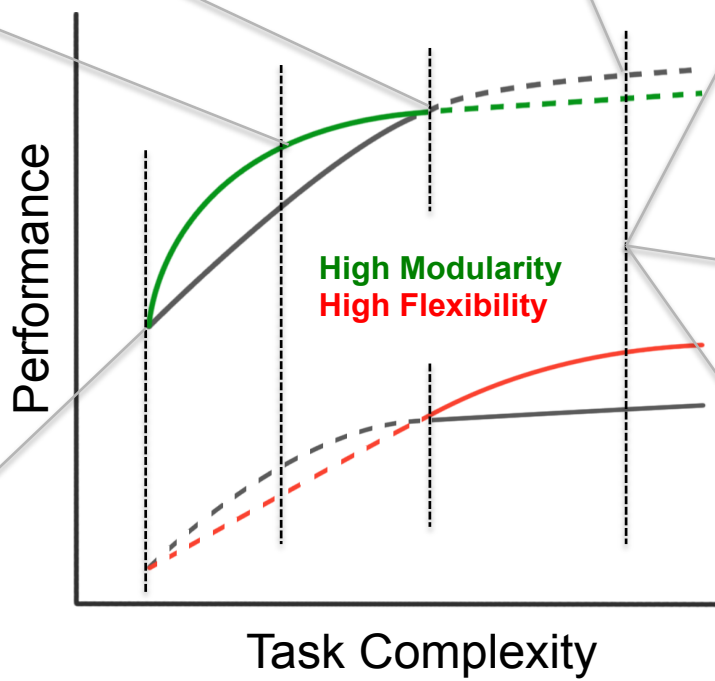

Performance

Task Complexity

**High Modularity**
**High Flexibility**

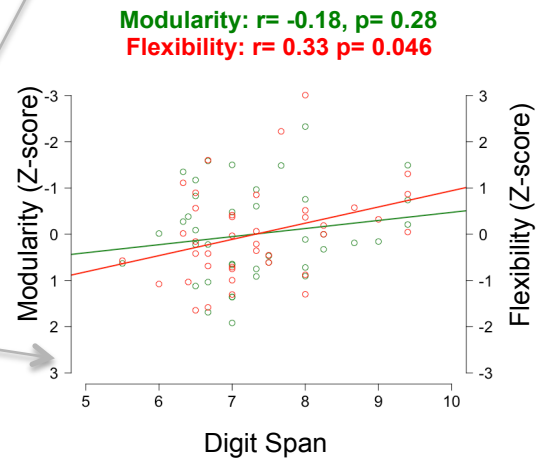

**Modularity: r= -0.18, p= 0.28**
**Flexibility: r= 0.33 p= 0.046**

Digit Span

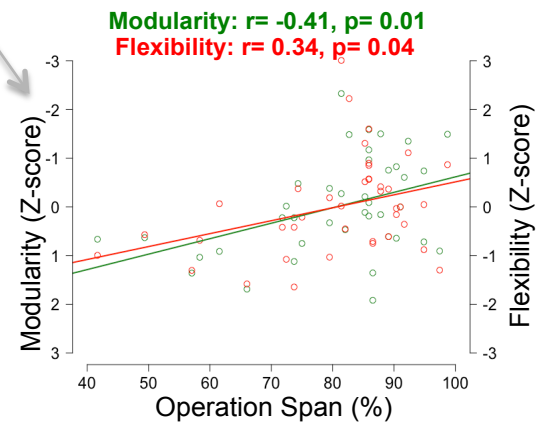

**Modularity: r= -0.41, p= 0.01**
**Flexibility: r= 0.34, p= 0.04**

Operation Span (%)

## BA Anatomical Parcellation Atlas

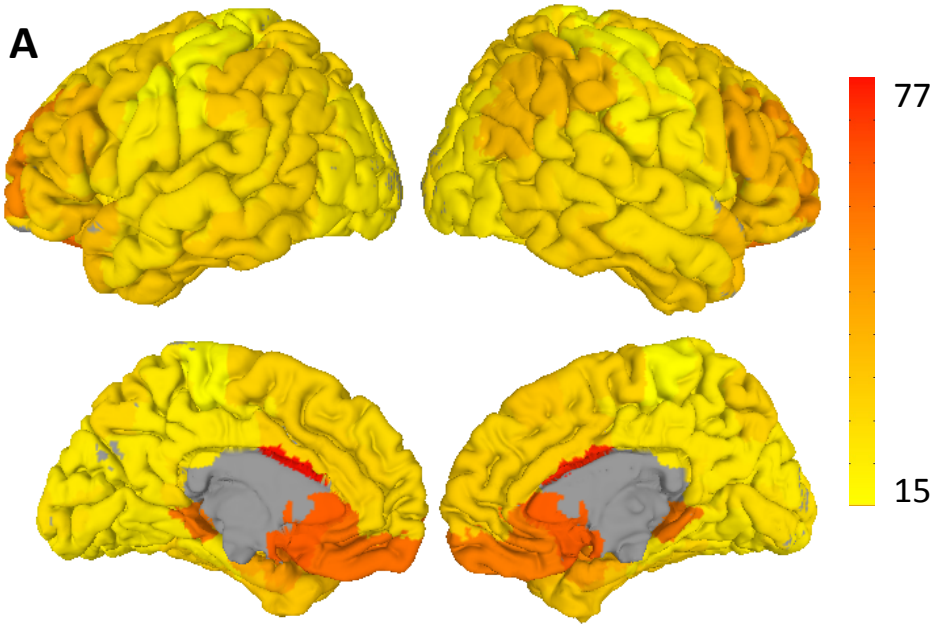

## Craddock 2012 Functional Parcellation Atlas

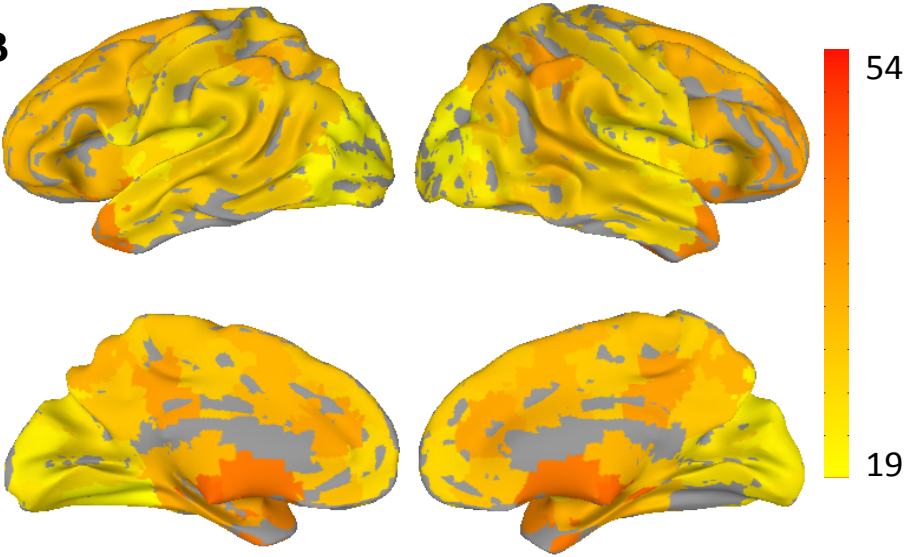

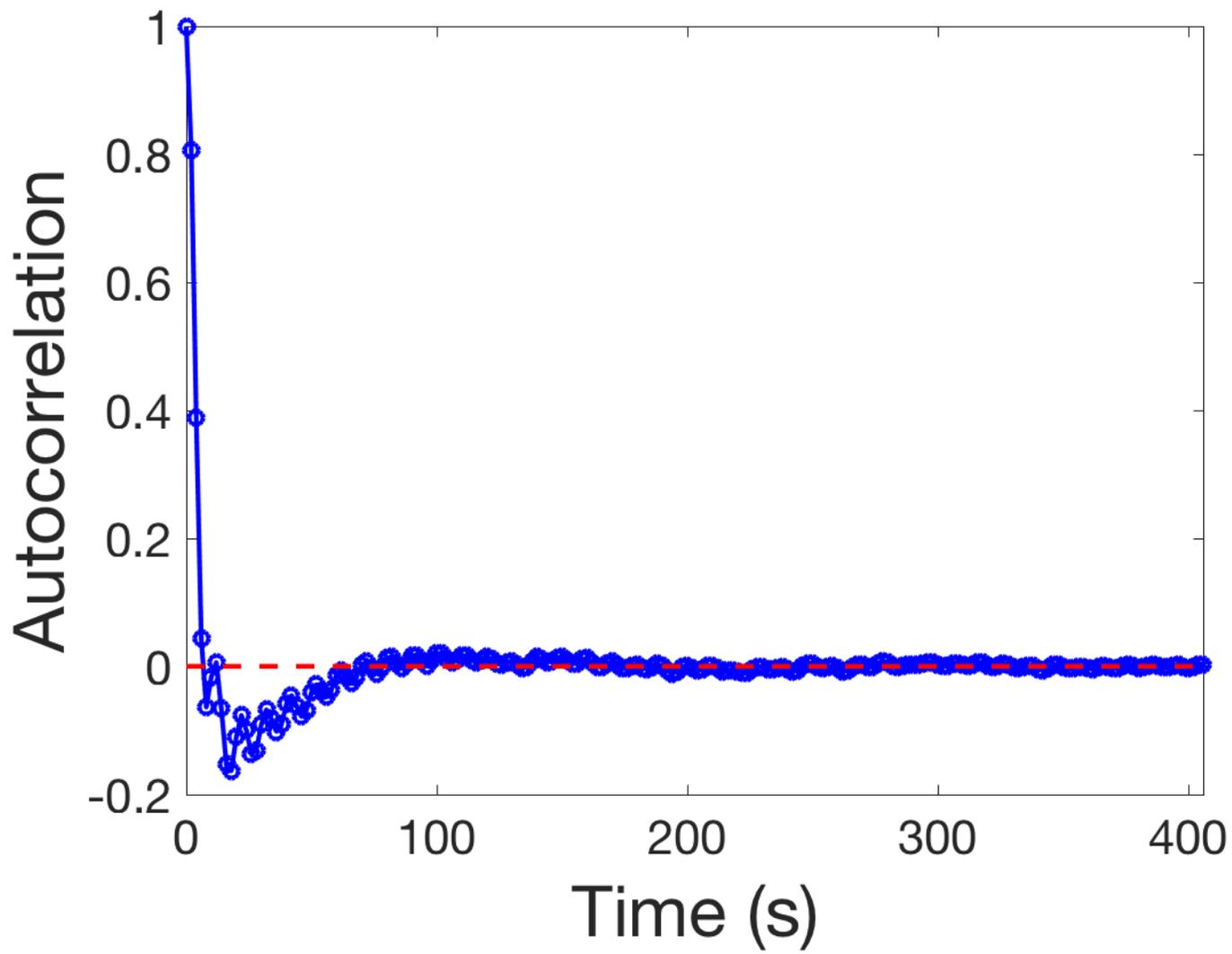